\documentclass[twocolumn,showpacs,preprintnumbers,amsmath,amssymb,prl]{revtex4}
\usepackage{graphicx}
\usepackage{dcolumn}
\usepackage{bm}
\begin{document}

\title{Nodeless $d$-wave superconducting pairing
due to residual antiferromagnetism in underdoped
Pr$_{2-x}$Ce$_x$CuO$_{4-\delta}$}

\author{Tanmoy Das, R. S. Markiewicz, and A. Bansil}
\address{Physics Department, Northeastern University, Boston MA 02115, USA}
\date{\today}
\begin{abstract}

We have investigated the doping dependence of the penetration
depth vs. temperature in electron doped
Pr$_{2-x}$Ce$_x$CuO$_{4-\delta}$ using a model which assumes the
uniform coexistence of (mean-field) antiferromagnetism and
superconductivity. Despite the presence of a $d_{x^2-y^2}$ pairing
gap in the underlying spectrum, we find nodeless behavior of the
low-$T$ penetration depth in underdoped case, in accord with
experimental results. As doping increases, a linear-in-$T$ behavior
of the penetration depth, characteristic of $d$-wave pairing,
emerges as the lower magnetic band crosses the Fermi level and
creates a nodal Fermi surface pocket.

\end{abstract}


\maketitle \narrowtext

An understanding of the symmetry of the order parameter and its
evolution with hole and electron doping is a key to unraveling the
mechanism of high-$T_c$ superconductivity in the cuprates. Many
experimental and theoretical studies of these fascinating
materials demonstrate the presence of antiferromagnetic (AFM)
order in underdoping for both hole-\cite{hosseini} and
electron doping\cite{plcco,kusko,tanmoy}. With hole-doping, the
route followed by the AFM phase as it develops into the
superconducting (SC) phase involves the intervention of nanoscale
phase separations related to stripe or pseudogap
physics\cite{hosseini}. The behavior with electron doping, on the
other hand, seems to be simpler in that the doped phase appears to
be a uniform AFM metal, possibly evolving into a phase with
coexisting AFM and SC orders\cite{plcco,kusko,tanmoy}.

For the hole doped cuprates it is generally believed that $d$-wave
pairing survives up to the edge of antiferromagnetism\cite{hosseini,dagan,tsuei}, but the doping
dependence of the pairing symmetry with electron doping remains a
matter of debate. This symmetry has been studied by low-$T$
penetration depth (PD)
measurements\cite{wu,andreone,schneider,alff,skinta,snezhko}, point
contact spectroscopy\cite{biswas,qazilbash},
tunneling\cite{chesca}, and other phase sensitive probes\cite{ariando},
in a variety of electron-doped cuprates, including
Nd$_{2-x}$Ce$_{x}$CuO$_{4-\delta}$
(NCCO)\cite{wu,andreone,schneider,alff,ariando},
La$_{2-x}$Ce$_{x}$CuO$_{4-\delta}$ (LCCO)\cite{skinta,chesca}
Pr$_{2-x}$Ce$_{x}$CuO$_{4-\delta}$
(PCCO)\cite{skinta,biswas,qazilbash,snezhko}.
The results have been contradictory, with some early
measurements\cite{wu,andreone,schneider,alff} finding evidence for
$s$-wave pairing, while other experiments suggest a transition from
$d$-wave in underdoping to either $s$-wave\cite{skinta,biswas}
or $(d+is)$-wave character\cite{qazilbash} in the optimally and
overdoped cases. Yet other experiments\cite{snezhko,ariando} report
only $d$-wave pairing, with the situation further complicated by the
presence of nonmonotonic SC-gap variations observed in
NCCO\cite{raman} and Pr$_{1-x}$LaCe$_{x}$CuO$_{4-\delta}$
(PLCCO)\cite{plcco}.

A recent study approximated the AFM background by treating the
resulting partially-gapped Fermi surface (FS) in a two band
model\cite{QHWang}. To understand the interplay between AFM and SC orders and
the role of AFM order in modifying the pairing\cite{voo,yuan},
in this article we directly evaluate the PD in a
model with coexisting AFM and SC order.  We
assume a SC gap of $d$-wave pairing with a combination of
first and third harmonics, which is necessary to incorporate
nonmonotonic gap variations\cite{raman,plcco}.  We find that even in
the presence of a $d$-wave
pairing gap, the PD varies exponentially at low $T$
for most dopings $-$ a behavior characteristic of a nodeless SC-gap,
as antiferromagnetism suppresses the spectral weight from the
nodal point. In the overdoped case ($x= 0.152$), the PD
shows a linear-in-$T$ behavior as the hole pocket forms in
the nodal region. Our analysis indicates that with increasing
electron doping the position of the maximum leading edge gap on
the FS moves away from the antinodal direction and that the
nonmonotonic nature of the gap becomes stronger.

Our treatment of the in-plane PD is based on the Hamiltonian
\begin{equation}
\label{hamiltonian} H = H_{pair} + H_{int},
\end{equation}
where $H_{pair}$ describes the physics of coexisting AFM and SC
orders. We take $H_{pair}$ to be a one-band, tight-binding Hubbard
Hamiltonian along the lines of Ref. \onlinecite{tanmoy} in which
the SC gap is of $d$-wave pairing with a combination of
first and third harmonics.  The tight binding
parameters are assumed to be same as for NCCO\cite{kusko}.
 The external perturbation is given by the electromagnetic
interaction,
 \begin{equation}\label{interaction}
 H_{int} =
-\left(\frac{e}{c}\right)\vec{A}\cdot\Bigl[\sum_{\vec{k},\sigma}\vec{v}_{
\vec{k}} c^{\dag}_{\vec{k},\sigma} c_{\vec{k},\sigma}\Bigr]
\end{equation}
where $c^{\dag}_{\vec{k},\sigma}$ ($c_{\vec{k},\sigma}$) is the
electronic creation (destruction) operator with momentum
$\vec{k}$, charge $e$ and spin $\sigma$, and $c$ is speed of
light. $\vec{A}$ is the Fourier component of the vector potential
in momentum space. $\vec{v}_{\vec{k}} =
\partial \xi_{\vec{k}}/(\hbar\partial \vec{k})$ is the band velocity
for the noninteracting band $\xi_{\vec{k}}$\cite{kusko}.

The PD is obtained by evaluating the induced
current parallel to the vector potential, which is proportional to
the inverse square of the in-plane PD\cite{tinkham}. We
have generalized the pure BCS result
to the mixed AFM-SC case and find\cite{qiang}
\begin{widetext}
\begin{eqnarray}\label{penetrationdepth}
\lambda_{ij}^{-2}(T)&=&\frac{4\pi e^2}{c^2a^2d}
\sum_{\nu=\pm}\sum_{\vec{k}}^{\prime}\left[\left(\frac{1}{m^{\nu}_{\vec{k}ij
}}\right) \left(1-\frac{\xi_{\vec{k}}^+ +\nu
E_{0\vec{k}}}{E_{\vec{k}}^{\nu}} \tanh{(\beta
E_{\vec{k}}^{\nu}/2)}\right)
-\frac{\beta}{2}v^{\nu}_{\vec{k}i}v^{\nu}_{\vec{k}j} {\rm
sech}^2(\beta E_{\vec{k}}^{\nu}/2)\right].
\end{eqnarray}
\end{widetext}
Here, $a$ is the in-plane and $d$ the out-of-plane lattice
constant of PCCO and $\beta = 1/k_BT$.  The prime on
the ${\vec k}$ summation means that the sum is restricted to wave
vectors in the magnetic zone. The magnetic field is assumed to lie
perpendicular to the CuO$_2$ plane. For a tetragonal lattice $\lambda_{ij}$
is diagonal, with $\lambda_{aa}=\lambda_{bb}=\lambda$ within the CuO$_2$ plane.
Interestingly, Eq.~3 displays a form similar to that for
a pure $d$-wave SC\cite{susumu,sheehy}, excepting two
modifications.
Firstly, the FS has components $\nu = \pm$ associated with the upper
magnetic band (UMB) and the lower magnetic band (LMB), respectively:
\begin{equation}\label{E}
(E_{\vec{k}}^{\nu})^2 = \Big(\xi_{\vec{k}}^{+}+\nu E_{0\vec{k}}
\Big)^2 + \Delta_{\vec{k}}^2,
\end{equation}
where $E_{0\vec{k}}=\sqrt{(\xi_{\vec{k}}^{-})^2 + (U_QS)^2}$ and
$\xi_{\vec{k}}^{\pm} = (\xi_{\vec{k}} \pm
\xi_{\vec{k}+\vec{Q}})/2$.
$\Delta_{\vec{k}}$ is the SC gap and $U_QS$ the AFM gap in terms of the
AFM repulsion $U_Q$ and the commensurate magnetisation $S$ at the nesting
vector
$Q=(\pi,\pi)$. Secondly, the band masses $m^{\nu}_{\vec{k}ij}$ and quasiparticle
velocities $\vec{v}^{\nu}_{\vec{k}}$ have magnetic
correlation corrections:
 $1/m^{\nu}_{\vec{k}ij}=\partial^2(\xi_{\vec{k}}^+
+\nu E_{0\vec{k}})/(\hbar^2\partial k_i\partial k_j)$,
$\vec{v}^{\nu}_{\vec{k}}=\partial(\xi_{\vec{k}}^{+} + \nu
E_{0\vec{k}})/(\hbar\partial \vec{k})$.

We obtain the AFM and SC gaps self-consistently as a function of
$T$ for a series of doping levels over the range $x= 0.115-0.152$,
using doping dependent interaction parameters\cite{tanmoy}, before
proceeding with the PD calculation from
Eq.~\ref{penetrationdepth}. The effective AFM interaction given by $U_Q$
is taken from our earlier work on NCCO\cite{kusko} and decreases
from a value of 3.33$t$ at $x$=0.115 to 3.1$t$ at $x$=0.152; the resulting
self-consistent magnetization $S$ decreases linearly from 0.2 to 0.13 over
this doping range, in agreement with earlier results\cite{kusko},
despite the presence of the SC order.

The two terms on the right hand side of Eq.
{\ref{penetrationdepth}} correspond to the conventional
diamagnetic (first term) and paramagnetic\cite{cooper} (second term)
currents of electrons.  In a London picture,
$\lambda^{-2}(T)$ is proportional to the SC electron
density $n_s$, and hence vanishes as $T\rightarrow T_c$, while at $T =0$
all the electrons are superconducting.  Here we find a similar result, but {\it only the electrons
in the AFM pockets condense.}
Similarly, as $T\rightarrow 0 $, the
linear-in-$T$ PD found in overdoped samples reveals
the presence of gap nodes, where normal quasiparticles persist to
zero energy.

\begin{figure}[tp]
\rotatebox{270}{\scalebox{0.27}{\includegraphics{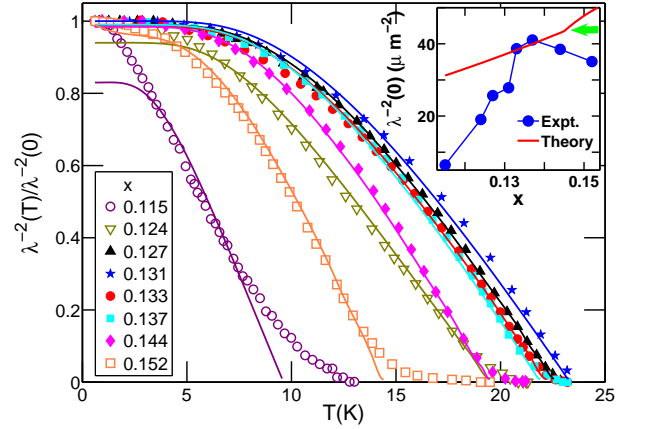}}}
\caption{(color online) Various colored lines give computed
$\lambda^{-2}(T)/\lambda^{-2}(0)$ as a function of $T$ for different dopings $x$; the corresponding
experimental data for PCCO\cite{pcco} is shown by symbols of the
same color (see legend). Inset: Computed (red line) and
experimental (blue dots) values of $\lambda^{-2}(0)$ as
a function of doping. Green arrow points to the {\it kink} associated with
the opening of the nodal pocket in the theory curve.}
\label{penetration}
\end{figure}

Figure~\ref{penetration} compares the theoretical and experimental values
of the inverse square of the PD in PCCO\cite{pcco} over the
doping range $x = 0.115-0.152$. The results are normalized to the computed
$T=0$ value for the $c$-axis lattice constant $d= 12.2 $ \AA\cite{markert}
in order to highlight $T$-dependencies. The overall agreement is
quite good, allowing us to adduce the doping dependence of the AFM and SC
properties as discussed below. A discrepancy is found at the lowest and
highest dopings, where the PD shows a tail extending beyond
$T_c$, possibly associated with sample inhomogeneities\cite{twoband}.

Turning to the inset in Fig.~1, note first that the theoretical values
(red line) of $\lambda^{-2}(0)\propto n_s(0)$, do not involve any
further fitting parameters beyond those used in fitting the $T$-dependence
of $\lambda$. Around optimal doping, theory and experiment are seen to
be in accord indicating that the theory correctly predicts the value of
$n_s$(0), although striking deviations are seen away from optimal doping.
Insight into this behavior is obtained by observing that as $T\rightarrow$
0, all the electrons on the FS condense so that $n_s$(0) is proportional to
the area of the FS pockets. For this reason, the computed $n_s$(0) decreases
linearly with underdoping and undergoes a change in slope (marked by the
green arrow) as the ($\pi$/2,$\pi$/2) pocket crosses the Fermi level in overdoping.
 In sharp contrast, the experimental points (blue dots)
present a peak around optimal doping and a loss of carriers away from
optimal doping, indicative of 'bad metal' physics where the
SC transition is dominated by thermal phase
fluctuations\cite{carlson,foot3}. These results suggest that on the
underdoped side AFM fluctuations are more deleterious than expected from
the mean field BCS model underlying our computations. The reappearance of
bad metal behavior on overdoping is puzzling and its origin is
unclear$-$it may be related to increasing doping-induced disorder.

\begin{figure}
\rotatebox{270}{\scalebox{0.33}{\includegraphics{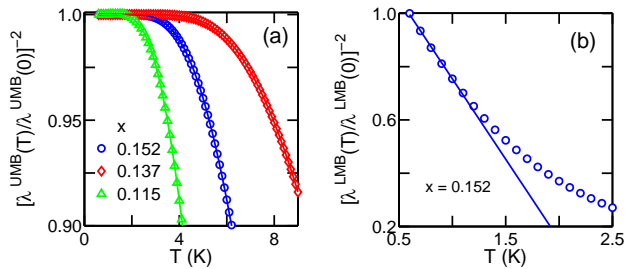}}}
\caption{(color online)(a) UMB contribution to the theoretical
PD at three different dopings. Solid lines give the
corresponding exponential fits of form $[1-C\exp{(-\beta D)}]$ at
low $T$. (b) LMB contribution at $x=0.152$ (circles), and the
related linear fit, $(1-aT)$, at low $T$ (blue line).}
 \label{penetfitting}
\end{figure}

\begin{figure}
\rotatebox{270}{\scalebox{0.33}{\includegraphics{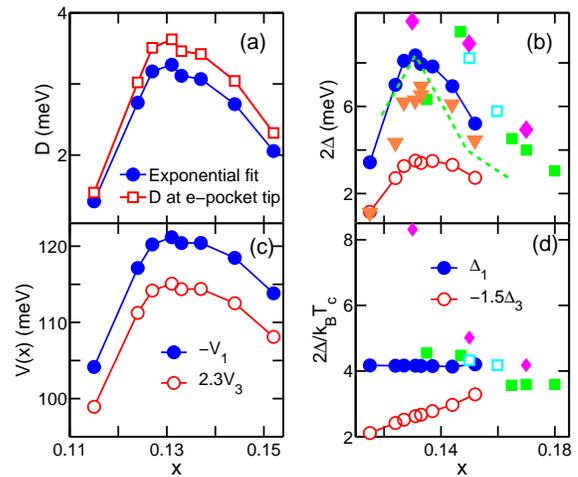}}}
\caption{(color online)
(a) SC-gap $D$ as a function of doping obtained from
exponential fits similar to those in Fig.~2(a) at low $T$ (blue
dots), and the values of the SC gap (red open squares) at the tip of the
electron pockets, shown in Fig.
\ref{fermisurfaces}(b)-(d)(yellow diamonds).
(b) Self-consistent first harmonic
$2\Delta_1$ (blue dots) and third harmonic $-4\Delta_3$
(red open circles) of the SC gap are compared with several
experimental results for NCCO and PCCO: Raman scattering in $B_{2g}$
channel from NCCO[PCCO]\cite{dqazilbash}, green filled [green open]
squares; Tunnelling data on NCCO\cite{biswas,qazilbash}, magenta diamonds; and two
band model computations\cite{QHWang}, orange triangles. Green dashed line
shows the scaled Raman $B_{2g}$ channel gap for NCCO.
(c) SC interaction potentials
as a function of doping: Absolute value of the $d$-wave first
harmonic $-V_1$ (blue dots) and the third harmonics $2.3V_3$
(red circles). (d) $2\Delta/k_BT_c$ for $\Delta_1$
and $-1.5\Delta_3$ for the first and third harmonic SC gaps are
compared to the experimental results. Various symbols have the
same meanings as in (b).}
\label{orderparameter}
\end{figure}

\begin{figure}
\rotatebox{0}{\scalebox{0.39}{\includegraphics{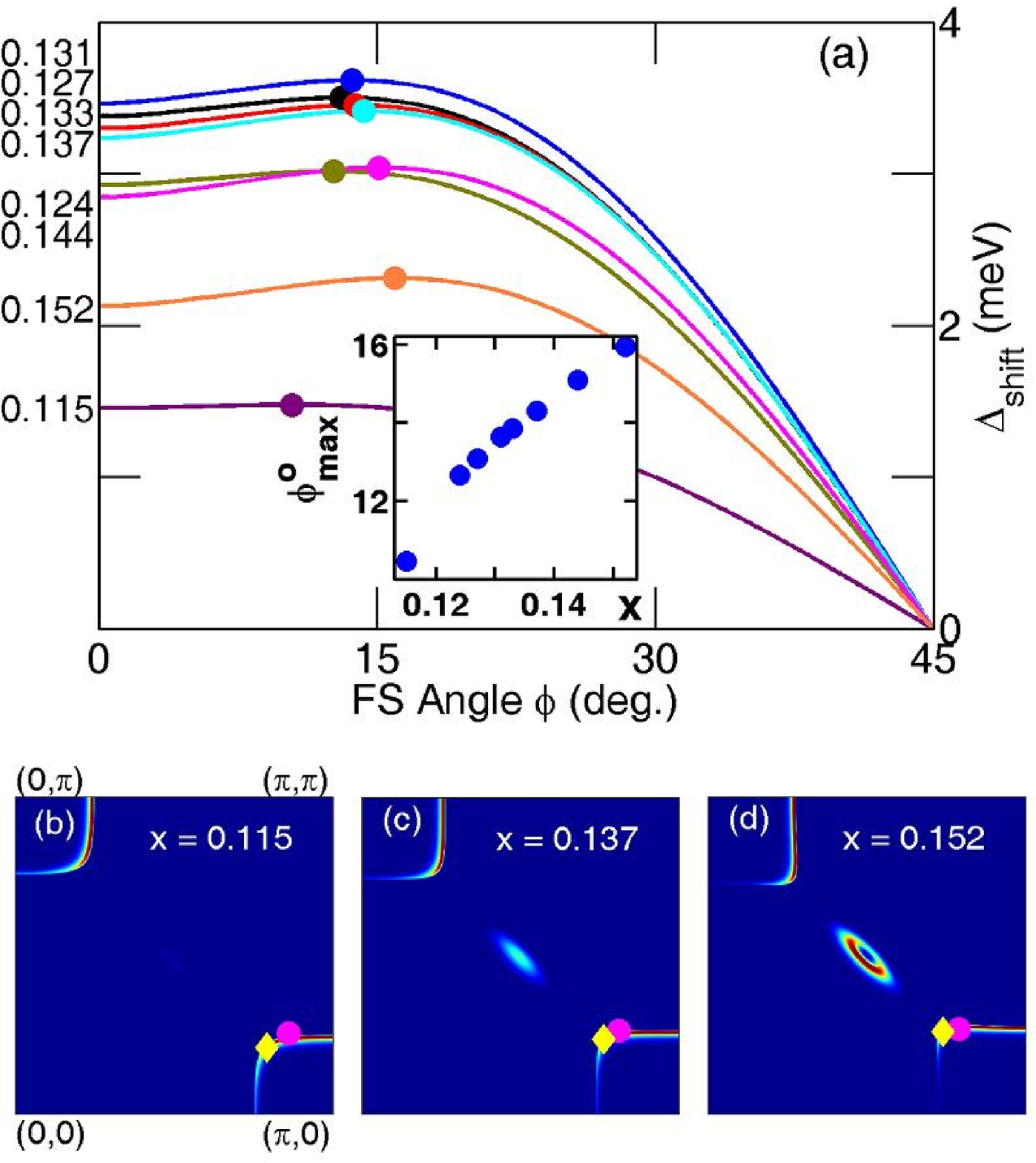}}}
\caption{(color online) a) Variation in the leading edge gap
(LEG), $\Delta_{\rm shift}$, over the FS in terms of the FS angle
$\phi$, where $\phi$ increases from zero along the antinodal
direction to 45$^o$ along the nodal direction. Solid dots mark
positions of the LEG maximum, $\phi_{\rm max}$, which is seen in
the inset to increase nearly linearly with doping. (b)-(d) FS
is calculated for three different dopings $x$. Red color denotes
maximum spectral weight and blue color zero intensity. Red dots give the position of the
maximum LEG gap on the FS considered in (a). Yellow diamonds mark
the tips of the UMB electron pockets, at which the SC gaps in Fig.
2(c) were calculated.} \label{fermisurfaces}
\end{figure}

We discuss the doping and $T$-dependencies of the preceding
theoretical PD results with reference to Figures
~\ref{penetfitting}-4.  In Fig. \ref{penetfitting}(a), we emphasize
that the contribution of the UMB is nodeless since the
PD is dominated by energies near the
Fermi level and the UMB pocket is far from the nodal region. An
exponential form, $1-C\exp{(-\beta D)}$, is seen to produce an excellent fit
in the low-$T$ region in Fig. \ref{penetfitting}(a). The values of the
SC-gap $D$ in Fig. \ref{orderparameter}(a) so obtained for the UMB (blue
dots) are quite close to the SC-gap (red open squares) at the electron
pocket tip, marked by yellow diamonds on the FS plots of
Figs.~\ref{fermisurfaces}(b)-(d). The value of $C$ is $\approx $4.4,
essentially independent of doping. Only at the highest doping $x=0.152$ do
we find a significant linear-in-$T$ ($d$-wave) contribution to the
PD as shown in Fig. \ref{penetfitting}(b), which coincides
with the appearance of the hole pocket near ($\pi$/2,$\pi$/2) at high
dopings as the LMB crosses the Fermi level (see Fig. 4(d)). A linear
equation of the form, $1-aT$, fits the LMB contribution very well up to $T$ =
1.5K as shown by blue line in Fig. \ref{penetfitting}(b)\cite{foot2}. We do not
find a second regime of linear-in-$T$ PD in the strongly
underdoped regime\cite{voo}.

Figure~\ref{orderparameter} examines the doping dependence of the SC gap
parameters. The dome-like shape as a function of doping of the gap $D$ in
Fig. \ref{orderparameter}(a) is reflected in the behaviors of the first
and third harmonics of the pairing gap in Fig.
\ref{orderparameter}(b) as well. Fig. 3(c) delineates the doping
dependence of the first and third harmonics of the pairing interaction,
which display a maximum near $x\approx$ 0.13 where $T_c$ is
optimal. The doping dependence of the SC gap parameters $\Delta_1$ and
$\Delta_3$ is compared with various experimental results in Figs. 3(b) and
(d). Some disagreement with Raman experiments on NCCO (green filled
squares) and PCCO (green open squares)\cite{dqazilbash} is due to sample
variations, reflected in $T_c$ variations, while
the ratio $2\Delta_1/k_BT_c$ is essentially constant and agrees well with
experiment. If we scale the experimental gap to fit the calculated maximum
at optimal doping, we can reproduce the dome-like behavior of the SC gap
as shown by the green dashed line in Fig \ref{orderparameter}(b) for NCCO.
The tunnelling data (magenta diamonds)\cite{biswas,qazilbash} do not
show a maximum, because tunnelling is sensitive to the {\it total gap}
obtained by combining AFM and SC gaps, and this combined gap in our
computations does not have a maximum near optimal doping. Similarly, the
larger gap seen by Raman\cite{dqazilbash} in under- and optimally doped
samples can be understood since the $B_{2g}$ channel measures the total
spectral gap near the ($\pi,0$) point\cite{liuraman}, and hence is
strongly coupled to the AFM order.

Fig.~\ref{orderparameter}(d) shows that the ratio $2\Delta_1/k_BT_c$ possesses a nearly constant value of 4.1, close
to the BCS value for a $d$-wave gap. In contrast, for
fixed ratio of $V_1/V_3$\cite{foot1}, the third harmonic ratio
$-2\Delta_3/k_BT_c$ increases linearly with doping. This is the
reason that the position of the maximum of the leading edge gap
(LEG) $\Delta_{\rm shift}$ on the FS, given by the FS angle
$\phi_{\rm max}$ in Fig. 4(a), moves away from the antinodal point
with doping; interestingly, the hot spots also move away with
doping from the antinodal direction, but their shift is much
smaller. We find that the ratio of the maximum value of
$\Delta_{\rm shift}$ to its value along the antinodal direction
increases with doping, indicating that the non-monotonic nature of
$d$-wave pairing symmetry becomes more pronounced as one goes from
under- to overdoping in the electron doped cuprates.

The evolution of the $(\pi/2,\pi/2)$-centered nodal hole-pocket is
seen in Figs.~\ref{fermisurfaces}(b)-(d). The absence of nodal pockets in
the underdoped regime (see (b)) is responsible for the nodeless
behavior of the SC gap. At optimal doping $x$=0.137, the hole pocket
is still $\sim25$ meV below $E_F$, but can be seen in (c) due to the
finite energy resolution. The nodal pocket is fully formed in the overdoped case of
(d) which is related to the striking $d$-wave behavior of PD
in Fig.~2(b) as well as the kink in $n_s$(0) in Fig.~1 inset.

In conclusion, we have shown that the linear-in-$T$ variation of
$\lambda^{-2}$ in electron doped cuprates is related to the appearance of
the $(\pi/2,\pi/2)$-nodal hole pocket on the FS, which occurs in the
overdoped regime. In underdoping, where the FS only consists of
the $(\pi,0)$-centered electron pockets, $\lambda^{-2}$ varies in a
nodeless manner, even though the pairing interaction is of $d$-wave
symmetry, because the electron pocket lies far from the nodal region. Our
analysis indicates that the SC electron density ($n_s(0)$) is
suppressed in a non-BCS fashion as one goes away from optimal doping to
either under- or overdoping. Interestingly, we find that
the SC interaction ($V_1$ and $V_3$) also peaks at optimal
doping.

\begin{acknowledgments}
This work is supported by the U.S.D.O.E contracts DE-FG02-07ER46352
and DE-AC03-76SF00098 and benefited from the allocation of supercomputer
time at NERSC and Northeastern University's Advanced Scientific Computation
Center (ASCC).
\end{acknowledgments}


\begin{thebibliography}{99}
\bibitem{twoband} The two band model\cite{QHWang} finds a different
interpretation for these tails, which we are unable to reproduce.
\bibitem{foot3} A very small $n_s$ may explain the deviations in the
$T$-dependence of PD for the lowest doped sample.
\bibitem{foot1} $T_c$ for each doping is
adjusted to produce the best agreement with the experimental
PD data and, in order to reduce the number of
parameters, the ratio between the SC interactions $V_1$ and $V_3$
has been kept fixed at a value of 2.4 for all dopings.
\end{thebibliography}
\end{document}